\documentclass[10pt]{article}  
\usepackage{graphicx}
\usepackage{amssymb}
\input epsf
\parskip=\medskipamount
\def\al{\alpha}

\def\k{\kappa}  
\def\kp{\kappa}  
   
\def\om{\omega}   
\def\IB{\relax{\rm l\kern-.18 em B}}
\def\IC{\relax{\rm l\kern-.50 em C}}
\def\IE{\relax{\rm l\kern-.12 em E}}
\def\IK{\relax{\rm l\kern-.18 em K}}
\def\IL{\relax{\rm I\kern-.18 em L}}
\def\IN{\relax{\rm I\kern-.18 em N}}
\def\IR{\relax{\rm I\kern-.18 em R}}
\def\smallonehalf{\frac{{}_1}{{}^2}}

\def\frac#1#2{{#1\over #2}}

\def\ptos{\leaders\hbox to 2mm{\hfil{.}\hfil}\hfill}
\def\\{\hfill\break}

\def\<#1>{\langle#1\rangle}

\def\Cos{\mathop{\rm C}\nolimits}    % funcion Coseno
\def\Sin{\mathop{\rm S}\nolimits}    % funcion Seno
\def\Tan{\mathop{\rm T}\nolimits}    % funcion Tangente
\def\k{\kappa}                       % kappa
\font\tenfrak=eufm10  \font\sevenfrak=eufm7  \font\fivefrak=eufm5
\newfam\frakfam
\textfont\frakfam=\tenfrak\scriptfont\frakfam=\sevenfrak
\scriptscriptfont\frakfam=\fivefrak

\font\tengoth=eufm10 scaled\magstep1 \font\sevengoth=eufm7
\font\fivegoth=eufm5
\newfam\gothfam
\textfont\gothfam=\tengoth\scriptfont\gothfam=\sevengoth
   \scriptscriptfont\gothfam=\fivegoth
\newtheorem{proposicion}{Proposition}

\begin{document}

\title{Superintegrability of the Post-Winternitz  system \\
on spherical and hyperbolic spaces}

\author{ Manuel F. Ra\~nada  \\ [3pt]
{\sl Dep. de F\'{\i}sica Te\'orica and IUMA } \\
  {\sl Universidad de Zaragoza, 50009 Zaragoza, Spain}     }
\date{  }
%-------------------------------------------------------------------
%%  Version vf  
%%  ArXiv Januart 2015 
%-------------------------------------------------------------------
\maketitle 
 
%----------------
%%  \begin{quote}
%%  {\tt  [Filename: \jobname.tex] }
%%  \end{quote}
%----------------

\begin{abstract}
The properties of the Tremblay-Turbiner-Winternitz system (related to the harmonic oscillator) were recently studied on the two-dimensional spherical $S_{\kappa}^2$ ($\kappa>0$)  and hiperbolic  $H_{\kappa}^2$ ($\kappa<0$) spaces (J. Phys. A : Math.  Theor. 47, 165203, 2014). 
In particular, it was proved the higher-order superintegrability of the TTW system by making use of  (i) a curvature-dependent formalism,  and (ii) existence of a complex factorization for the additional constant of motion.   Now a similar study is presented for the Post-Winternitz  system (related to the Kepler problem). 
The curvature $\kappa$ is considered as a parameter and all the results are formulated in explicit dependence of $\kappa$. This technique leads to a correct definition of the Post-Winternitz (PW) system on spaces with curvature $\kappa$, to a proof of the existence of higher-order superintegrability (in both cases, $\kappa>0$ and $\kappa<0$),  and to the explicit expression of the  constants of motion.
\end{abstract}

\begin{quote}
%----------------
 Integrability on spaces of constant curvature. Superintegrability. Higher-order  constants of motion. Complex factorization. 

{\sl Running title:}{\enskip}
The PW system on spaces of constant curvature.

%----------------
%%  37J35   Completely integrable systems, topological structure
%%  70H06  Completely integrable systems and methods of integration

%----------------
%%  02.30.Ik   Integrable systems
%%  05.45.-a   Nonlinear dynamics and chaos
%%  45.20.Jj   Lagrangian and Hamiltonian mechanics
\end{quote}

\vfill
%----------------
\footnoterule{\small
\begin{quote}
  {\tt E-mail: {mfran@unizar.es}  }
\end{quote}
}

%----------------
\newpage
%----------------

%-----------------------------------------------------
%%  Section 1
\section{Introduction}

Let us denote by   $V_r$, $r=a,b,c,d$,  the four two-dimensional potentials whith separability in two different coordinate systems in the Euclidean plane; 
they are therefore  superintegrable with quadratic constants of motion (see \cite{MiPWJPa13} for a recent review on superintegrability). 
We now consider two of them; $V_a$ that is separable in Cartesian and polar coordinates
\begin{equation}
   V_{a}  =  {\smallonehalf}\om_0^2(x^2 + y^2) + \frac{k_2}{x^2} + \frac{k_3}{y^2} \,,
\end{equation}
and $V_c$ that is separable in polar and parabolic coordinates
\begin{equation}
   V_{c}  =  \frac{k_1}{\sqrt{x^2 + y^2}} + \frac{k_2}{y^2} + \frac{k_3\,x}{y^2 \sqrt{x^2 + y^2}}   \,. 
\end{equation}
The potential $V_a$ (related to the harmonic oscillator) admits two superintegrable generalizations; they are separable but in only one coordinate system and because of this the third additional integral of motion  is not quadratic but a higher order polynomial.
 The first one
\begin{equation}
  V_{a}(n_x,n_y)  =  {\smallonehalf}{\om_0}^2(n_x^2 x^2 + n_y^2 y^2)
    + \frac{k_2}{x^2}  + \frac{k_3}{y^2}  \,,  \label{Van1n2}
\end{equation}
that preserves the separability in Cartesian coordinates,  was studied in \cite{RaSaMontreal}--\cite{RaRoS10}.  The second generalization, that takes the form 
\begin{equation}
  V_{ttw}(r,\phi)  =  {\smallonehalf}\,{\om_0}^2 r^2 +  \frac{1}{2\,r^2}\,\Bigl(   \frac{\alpha}{\cos^2(m\phi)}  +  \frac{\beta}{\sin^2(m\phi)} \Bigr) \,,   
\label{VttwE2}
\end{equation}
was firstly  studied by Tremblay, Turbiner, and Winternitz \cite{TTW09}--\cite{TTW10}, and then by other authors \cite{Qu10a}--\cite{Ra14JPa}. 
When $m=1$ it reduces to $V_{a}$, but in the general $m\ne 1$ case 
($m$ must be an integer or rational number)
it  is only separable in polar coordinates; therefore, the third integral is not quadratic in the momenta but a polynomial  of higher order than two (the degree of the polynomial depends of the value of $m$).

The potential $V_c$ (related to the Kepler problem), that  in polar coordinates becomes 
\begin{equation}
   V_{c}  =  -\, \frac{g}{r}  +  \frac{F_1(\phi)}{r^2} \,,{\quad}
  F_1(\phi) =  \frac{k_2 }{\sin^2{\phi}}  + \frac{k_3 \cos\phi }{\sin^2{\phi}}   \,,
  \label{VcF1}
\end{equation}
can also be written as follows 
$$
   V_{c}  =  -\, \frac{g}{r}   +   \frac{1}{r^2}\Bigl(\frac{\alpha} {\cos^2(\phi/2)} +  \frac {\beta} {\sin^2(\phi/2)}\Bigr)   \,,  \ 
   k_2 = 2(\al+\beta)\,, \ k_3=2(\beta-\al) \,,  
$$
Therefore, the angular-dependent functions in the potentials $V_a$ and $V_c$ appear as two particular cases, $m=1$ and $m=1/2$, of a more general angular function.  
It seems therefore natural to suppose that the existence of  $V_{ttw}$,  considered as a generalization of the potential $V_a$, must determine a similar generaliza\-tion of the potential $V_c$. 
In fact $V_{c}$ also admits a superintegrable generalization of higher order 
\begin{equation}
  V_{pw}(r,\phi)  =  -\, \frac{g}{r}  +  \frac{1}{2\,r^2}\,\Bigl(   \frac{\alpha}{\cos^2(m\phi)} + \frac{\beta}{\sin^2(m\phi)} \Bigr) \,,   \label{VpwE2}
\end{equation}
that is quite similar to the $V_{ttw}$ potential. 
The first proof of the superintegrability of this new potential was given in \cite{PostWint10} by relating $V_{pw}$  with $V_{ttw}$ via a coupling constant  metamorphosis transformation \cite{KaMiPost10}. 

The following two points summarize some properties and references important for this paper.
%----------------
\begin{itemize}

\item
The superintegrability  of the potential  $V_{a}(n_x,n_y)$ was proved making use of two different approaches, dimensional reduction  \cite{EvVe08} --\cite{RodTW09} and complex factorization \cite {RaRoS10}. The idea of the method presented in \cite {RaRoS10} is that the third additional integral  can be obtained as the product of powers of two simple complex functions. 
More recently a very similar method was applied to prove the superintegrability of the TTW  \cite{Ra12JPaTTW} and the PW \cite{Ra13JPaPW} systems. 

\item
The TTW system has been recently studied on spaces of constant curvature $\kappa$ using different approaches \cite{ChDgR11, Hakob12PLa, GonKas14AnnPhys,Ra14JPa}.  In fact, it was proved in \cite{Ra14JPa} that it can be correctly defined in the two cases, that is, spherical $\kappa>0$ and hyperbolic $\kappa<0$, and that the curved version of the TTW system is superintegrable as well.  An important point is that the expression of the third additional integral of motion was also  explicitly obtained by making use of the complex factorization approach.  

\end{itemize}

The main purpose of this paper is to study the PW system on spaces of constant curvature;  the idea is to translate the results obtained in \cite{Ra13JPaPW} in the Euclidean space  to spherical and hyperbolic spaces. 
In fact the structure of this article is very similar to that of \cite{Ra14JPa}  and we also use the same notation, the same formalism (curvature-dependent formalism)  and the same approach (existence of appropriate complex functions).

%-----------------------------------------------------
%%  Section 2
\section{The Kepler and the Kepler-related  potential $V_c$ on spaces of constant curvature }

%-----------------------------------------------------
%%   Seccion 2.1
\subsection{Notation and curvature-dependent formalism  }

The function $F_1(\phi)$ in the expression (\ref{VcF1}) of $V_{c}$ is just the particular case $m=1$ of the following angular function 
\begin{equation}
   F_m(\phi) = \frac{k_a} {\sin^2(m\phi)} +  k_b\,\Bigl(\frac {\cos(m\phi)} {\sin^2(m\phi)}\Bigr) \,,  \label{Fm}
\end{equation}
that is related with the angular functions in the original expressions of the TTW and PW potentials by the following trigonometric equality 
\begin{equation}
  \frac{2(\alpha+\beta)} {\sin^2(2m\phi)} +  2(\beta - \alpha)\,\Bigl(\frac {\cos(2 m\phi)} {\sin^2(2m\phi)}\Bigr) 
  =  \frac{\alpha} {\cos^2(m\phi)} +  \frac {\beta} {\sin^2(m\phi)}  \,.   
\label{TrigEquality}
\end{equation}
The two particular cases, $m=1$ and $m=2$, that take the form 
\begin{eqnarray*}
 \frac{F_{1}(\phi)}{r^2} &=& \frac{k_a}{y^2} + \frac{k_b\,x}{y^2\sqrt{x^2+y^2}}  
 \,,{\quad} (m=1)  \cr
 \frac{F_{2}(\phi)}{r^2} &=& \frac{k_a-k_b}{4x^2} + \frac{k_a+k_b}{4y^2}  
  \,,{\quad} (m=2)  
\end{eqnarray*}
are just the angular-dependent functions in the potentials $V_c$ and $V_a$. 

 In what follows we consider  the curvature $\k$ as a parameter and in order to obtain equations valid for the three possible values of $\kappa$ (that is, $\kp>0$, $\kp=0$, or $\kp<0$), we introduce the following trigonometric-hyperbolic functions $\Cos_{\kp}(x) $ and $\Sin_{\kp}(x) $ defined as 
 $$
 \Cos_{\kp}(x) =\cos{\sqrt{\kp}\,x}  \,,{\quad} 
 \Sin_{\kp}(x) =\frac{1}{\sqrt{\kp}} \sin{\sqrt{\kp}\,x}  \,,{\quad} \kp\in\IR\,,
 $$
that can be rewritten with more detail as follows 
%----------------
\begin{equation}
  \Cos_{\kp}(x) = \cases{
  \cos{\sqrt{\kp}\,x}       &if $\kp>0$, \cr
  {\quad}  1               &if $\kp=0$, \cr
  \cosh\!{\sqrt{-\kp}\,x}   &if $\kp<0$, \cr}{\qquad}
%----------------
  \Sin_{\kp}(x) = \cases{
  \frac{1}{\sqrt{\kp}} \sin{\sqrt{\kp}\,x}     &if $\kp>0$, \cr
  {\quad}   x                                &if $\kp=0$, \cr
  \frac{1}{\sqrt{-\kp}}\sinh\!{\sqrt{-\kp}\,x} &if $\kp<0$, \label{SkCk}\cr}
\end{equation}
%----------------
and $ \Tan_{\kp}(x) = \Sin_{\kp}(x)/\Cos_{\kp}(x)$ (more properties and references in \cite{Ra14JPa}). 
Then the expression, in geodesic polar coordinates $(r,\phi)$, of the
differential line element  on the spaces $S_{\k}^2$ ($\kp>0$), $\IE^2$ ($\kp=0$), and $H_{\kp}^2$ ($\kp<0$), with constant curvature $\kp$ is given by 
\begin{equation}
 ds_\kp^2 = d r^2 + \Sin_{\kp}^2(r)\,d{\phi}^2 \,,
\end{equation}
so that it reduces to
$$
 ds_1^2  =  d r^2 + (\sin^2 r)\,d{\phi}^2 \,,{\quad}
 ds_0^2  =  d r^2 + r^2\,d{\phi}^2 \,,{\quad}
 ds_{-1}^2 = d r^2 + (\sinh^2 r)\,d{\phi}^2\,,
$$
in the three particular cases of the unit sphere $S_{1}^2$ ($\kp=1$), Euclidean plane $\IE^2$ ($\kp=0$), and `unit' Lobachewski plane $H_{-1}^2$ ($\kp=-1$).  

The kinetic term (Lagrangian of the geodesic motion) has the following form
$$
  T = {\smallonehalf}\,\Bigl(\,v_r^2 + \Sin_{\kp}^2(r) v_{\phi}^2\,\Bigr)  \,,
$$
the Noether momenta, reducing to the two linear momenta $p_x$ and $p_y$ in the Euclidean case, are given by 
\begin{eqnarray*}
 P_1(\kp)  &=&  (\cos{\phi})\,v_r - (\Cos_{\kp}(r) \Sin_{\kp}(r)\sin{\phi})\,v_{\phi}  \,,\cr
 P_2(\kp)  &=& (\sin{\phi})\,v_r + (\Cos_{\kp}(r) \Sin_{\kp}(r)\cos{\phi})\,v_{\phi} \,,
\end{eqnarray*}
and the $\k$-dependent angular momentum has the following expression  
$$
 J(\kp)  = \Sin_{\kp}^2(r)\,v_{\phi} \,.
$$

%-----------------------------------------------------
%%   Seccion 2.2
\subsection{The Kepler  potential on spaces of constant curvature } 

 The following (spherical, Euclidean, hyperbolic) Lagrangian with curvature $\kp$,
\begin{equation}
 L(\kp) = {\smallonehalf}\,\Bigl(\,v_r^2 + \Sin_{\kp}^2(r) v_{\phi}^2\,\Bigr)
    -  U(r;\kp) \,,{\quad}
   U(r;\kp) = -\,\frac{g}{ \Tan_{\kp}(r)} \,,
\end{equation}
represents the $\kp$-dependent version of the Kepler problem in such a way that the potential $U(r;\kp)$ reduces to
$$
 U_1 =  -\,\frac{g}{\tan r}  \,,{\quad}
 U_0 = V =  -\,\frac{g}{r}  \,,{\quad}
 U_{-1} = -\,\frac{g}{\tanh r}  \,,
 $$
in the three particular cases of the unit sphere ($\kp=1$), Euclidean plane ($\kp=0$),  and `unit' Lobachewski plane ($\kp=-1$) \cite{CRS05Jmp}. The Euclidean function $V(r)$ appears in this formalism as making separation between two different behaviours (see Figure 1).  Of course, the domain of $r$ depends of the value of $\kp$; we have $r\in[0,\infty)$ for $\kp\le 0$ and $r\in[0,\pi /\sqrt{\kp}]$ for $\kp>0$. 
It is known  that this system is superintegrable for all the values of the curvature $\kp$ since that, in addition to the total Energy and the angular momentum $J(\k)$, it is endowed with the following two quadratic constants of the motion
%----------------
\begin{eqnarray*}
 I_3(\kp) &=&  P_2(\kp) J(\kp) -  g \cos\phi \,,\cr
 I_4(\kp) &=&  P_1(\kp) J(\kp) + g \sin\phi \,,  
\end{eqnarray*}
%----------------
that represent the  curvature versions of the two-dimensional Runge-Lenz
constant of motion, whose existence is a consequence of the additional
separability of $U(r;\kp)$ in two different systems of $\kp$-dependent
'parabolic' coordinates.

%-----------------------------------------------------
%%   Seccion 2.3
\subsection{The Kepler-related potential $V_c$ on spaces of constant curvature }

 The following (spherical, Euclidean, hyperbolic)  $\kp$-dependent potential 
\begin{equation}
 U_c(r,\phi;\kp) =  -\,\frac{g}{ \Tan_{\kp}(r)} 
 +   \frac{k_2 }{(\Sin_{\kp}(r)\sin{\phi})^2}  +\frac{k_3 \cos\phi }{(\Sin_{\kp}(r)\sin{\phi})^2}  \,,
\end{equation}
that is well defined for all the values of $\kp$, represents the spherical ($\kp>0$) and hyperbolic ($\kp<0$) version of the Euclidean potential $V_{c}$ ($\kp=0$);  it reduces to
\begin{eqnarray}
  U_c(r,\phi;1)   &=&    -\,\frac{g}{\tan r}   + 
 \frac{1}{\sin^2 r} \Bigl(  \frac{k_2 }{\sin^2{\phi}}  + \frac{k_3 \cos\phi}{\sin^2{\phi}} \Bigr)  \,,\cr
 U_c(r,\phi;-1)  &=&   -\,\frac{g}{\tanh r}   +
 \frac{1}{\sinh^2 r} \Bigl(  \frac{k_2 }{\sin^2{\phi}}  + \frac{k_3 \cos\phi}{\sin^2{\phi}} \Bigr)  \,,
{\nonumber}
\end{eqnarray}
in the particular cases of the unit sphere ($\kp=1$) and `unit' Lobachewski plane ($\kp=-1$). It is endowed with the following two quadratic constants of the motion
%----------------
\begin{eqnarray*} 
   I_2(\kp) &=& J^2(\kp) + \frac{2\,k_2 }{\sin^2{\phi}}  + \frac{2\,k_3\cos\phi }{\sin^2{\phi}}  \,,\cr 
  I_3(\kp) &=& P_2(\k)J(\kp)  - g \cos\phi + \frac{2 k_2}{ \Tan_{\kp}(r)} \Bigl( \frac{\cos\phi }{\sin^2{\phi} }  \Bigr)  + \frac{k_3 }{ \Tan_{\kp}(r)} \Bigl( \frac{1 + \cos^2\phi}{\sin^2{\phi} } \Bigr) \,,   
\end{eqnarray*}
%--------------- 
($I_1(\kp)$ is the energy) and  it is, therefore, a superintegrable system for all the values of the curvature $\kp$.

%-----------------------------------------------------
%%  Section 3  
\section{The PW system on spaces of constant curvature }

In the following, we will make use of the Hamiltonian formalism; therefore, the time 
derivative $d/dt$ of a function means the Poisson bracket of the function with the Hamiltonian.

We have seen, in the previous section 2, that in the two cases,  Kepler  and   $V_c$ potentials, the curvature $\kappa$ modify many characteristics but preserve the fundamental property of superintegrability. Now in this section we will prove that this is also true for the PW system

It is well known that  if $F(\phi)$ are arbitrary function then the following Hamiltonian  (Kepler plus an angular deformation introduced by $F(\phi$))
\begin{equation}
 H =   {\smallonehalf}\,\bigl(p_r^2 + \frac{p_\phi^2}{r^2}\bigr) -\,\frac{g}{r} 
  +  \frac{F(\phi)}{r^2} \,.  \label{H(rfi}
\end{equation}
is separable in  polar coordinates and  it is therefore Liouville integrable.

The following propsition states this property for spherical $S_\kp^2$ ($\kp>0$) and hyperbolic $H_{\kp}^2$ ($\kp<0$) spaces. 

%----------------
%%  (Proposicion 1)
\begin{proposicion}  \label{prop1}
The Hamiltonian
\begin{equation}
   H(\kp)  =  {\smallonehalf}\,\Bigl(\, p_r^2 +  \frac{p_{\phi}^2}{\Sin_{\kp}^2(r)}\,\Bigr) -\,\frac{g}{ \Tan_{\kp}(r)} +  \frac{F(\phi)}{(\Sin_{\kp}(r))^2}  \label{Hk}
\end{equation}
is separable in geodesic polar coordinates $(r,\phi)$ and  it is endowed with the following  two quadratic constants of the motion 
\begin{eqnarray*}  
 J_1(\kp) &=& p_r^2 + \frac{p_{\phi}^2}{\Sin_{\kp}^2(r)} -\,\frac{2g}{ \Tan_{\kp}(r)} + \frac{2F(\phi)}{(\Sin_{\kp}(r))^2}  \cr 
 J_2(\kp) &=& p_\phi^2 +2 F(\phi)
\end{eqnarray*}
 This property is true for all the values of the curvature $\kp$. 
\end{proposicion}

As we comment in the introduction, the PW system is separable in the Euclidean plane in polar coordinates. Now we see that it admits a generalization to the spaces $S_\kp^2$  ($\kp>0$) and $H_\kp^2$  ($\kp<0$) that appears as a particular case of the Hamiltonian (\ref{Hk}); therefore, it is also separable (and therefore integrable) in  spherical and hyperbolic spaces. 

The following proposition proves the superintegrability of the PW system on spaces of constant curvature $\kappa$ and presents a method for obtaining the explicit expression of the third integral of motion. 
%----------------
%%  (Proposicion 2)
\begin{proposicion}  \label{prop2}
Consider the noncentral Kepler-related potential
\begin{equation}
H_m(\kp)  =  {\smallonehalf}\,\Bigl(\, p_r^2 +  \frac{p_{\phi}^2}{\Sin_{\kp}^2(r)}\,\Bigr) + U_{m}(r,\phi) \,,{\quad}  
U_{m}(r,\phi)  =  -\,\frac{g}{ \Tan_{\kp}(r)} + \frac{F_m(\phi)}{(\Sin_{\kp}(r))^2}   \,,
\label{HPWk}
\end{equation}
where $F_m(\phi)$ is the following angular function 
$$
   F_m(\phi) = \frac{k_a} {\sin^2(m\phi)} +  k_b\,\Bigl(\frac {\cos(m\phi)} {\sin^2(m\phi)}\Bigr) \,,  \label{VkFm}
$$
and $k_a$ and $k_b$ are arbitrary constants.
Let $J_1$ and $J_2$ the two quadratic constants of motion associated to the Liouville integrability 
\begin{eqnarray*}  
 J_1(\k) &=& p_r^2 + \frac{p_{\phi}^2}{\Sin_{\kp}^2(r)} -\,\frac{2g}{ \Tan_{\kp}(r)}+ \frac{2 F_m(\phi)}{\Sin_{\kp}^2(r)}  \cr 
 J_2 &=& p_\phi^2 + 2 F_m(\phi)
\end{eqnarray*}
and let $M_r$ and $N_\phi$ be the  complex functions   $M_r = M_{r1} + i\, M_{r2}$ and  $N_{\phi} = N_{\phi 1} + i\, N_{\phi 2}$ with real and imaginary parts, $M_{r a}$ and $N_{\phi a}$, $a=1,2$, be defined as 
$$
 M_{r1} =  p_r\,\sqrt{J_2} \,,{\qquad}
 M_{r2} =  g -  \frac{J_2}{ \Tan_{\kp}(r)}   \,,
$$
$$
 N_{\phi 1} =   k_b  +  J_2\cos(m\phi)  \,,{\qquad}
 N_{\phi 2} =   p_\phi\,\sqrt{J_2}\, \sin(m\phi) \,.
$$ 
Then, the complex function $K_m$ defined as
$$
  K_m = M_r^{m} \,N_\phi^{*}
$$
is a (complex) constant of the motion.  
\end{proposicion}

{\it Proof:} 
First, let us comment that the expressions of the function $M_r$ is rather the same that in the Euclidean case \cite{Ra13JPaPW} but with the function $\Tan_\k(r)$ instead of $r$.  It satisfies the correct Euclidean limit 
 $$
  \lim_{\kp\to 0} M_{r}  =  \Bigl(p_r\,\sqrt{J_2}\Bigr) + i\, \Bigl(g -  \frac{J_2}{r} \Bigr)\,. 
$$
The expresions of the angular functions $N_{\phi 1}$ and $N_{\phi 2}$ are the same as in the Euclidean plane.

The time-derivative (Poisson bracket with $H(\kappa)$) of the function $M_{r1}$ is proportional to $M_{r2}$ and the time-derivative of the $M_{r2}$ is proportional to $M_{r1}$ but with the opposite sign 
$$
 \frac{d}{d t}\,M_{r1}=  -\, \lambda_{\kp}\,M_{r2} \,,{\quad}
 \frac{d}{d t}\,M_{r2} =   \lambda_{\kp}\,M_{r1}  \,,
$$
and this property is also true for the angular functions 
$$
 \frac{d}{d t}\,N_{\phi 1} =  -\,m\,\lambda_{\kp}\,N_{\phi 2} \,,{\quad}
 \frac{d}{d t}\,N_{\phi 2} =   m\,\lambda_{\kp}\,N_{\phi 1}  \,,{\quad} 
$$
where the common factor ${\lambda_\kp}$ takes the value 
$$
 {\lambda_\k} = \frac{1}{\Sin_{\kp}^2(r)}\,\sqrt{J_2}  \,,{\quad}
  {\lambda_0} = \frac{1}{r^2}\,\sqrt{J_2}   \,. 
$$
Therefore, the time-evolution of the complex functions $M_r$ and $N_\phi$ is given by
$$
 \frac{d}{d t}\,M_r  =  i\,   {\lambda_\kp}\,M_r  \,,{\quad}
 \frac{d}{d t}\,N_\phi  =   i\,  m\,{\lambda_\kp}\,N_\phi   \,,{\quad}
$$
Thus we have
%----------------
\begin{eqnarray*}
  \frac{d}{dt}\,K_m &=&  \frac{d}{dt}\,\Bigl( M_r^{m} \,N_\phi^{*} \Bigr) 
  = M_r^{(m-1)} \,\Bigl( \, m\,\dot{M_r}\,N_\phi^{*}   
  +   M_r\,\dot{N_\phi}^{*} \,\Bigr)   \cr
  &=& M_r^{(m-1)}\,\Bigl( \, m\, ( i\,   {\lambda_\kp}\,M_r)\,N_\phi^{*}   +   M_r\,(-\, i\,  m\,{\lambda_\kp}\,N_\phi^{*}) \,\Bigr) =  0  \,.
\end{eqnarray*}

Finally, let us comment that the moduli of these two complex functions, that are constant of the motion of fourth order in the momenta,  are given by
\begin{eqnarray*}
\mid M_r \mid^2  &=& (2 H - \kp J_2) J_2 + g^2 \cr
\mid N_\phi \mid^2 &=&  J_2^2 - 2 k_a J_2 +  k_b^2 
\end{eqnarray*}
\hfill${\square}$

Summarizing: the Hamiltonian $H_m(\kp)$ given by (\ref{HPWk}), that represents the  PW system on a space with curvature $\kappa$, is super-integrable for any value of the curvature (positive, zero or negative) and the additional constant of motion $K_m$ can be obtained by  complex factorization. 
Of course this property remains true when the  potential $U_{m}(r,\phi)$ is rewritten with the original way of presenting the angular-dependent function as in (\ref{VpwE2}) just by making use of the trigonometric equality (\ref{TrigEquality}). 

Since the function $K_m$ is complex it can be written as $K_m=J_3 + i\, J_4$ with $J_3$ and $J_4$ real constants of the motion, that is, $dJ_3/dt=0$ and $dJ_4/dt=0$. One of them, for example $J_3$, can be chosen as the third fundamental integral of the motion.

%-----------------------------------------------------
%%  Section 4   
\section{Final comments } 

  In the previous paper \cite{Ra14JPa} we studied the TTW system and we proved that it was not a specific characteristic of the Euclidean space but it is well defined as a superintegrable system in all the three spaces of constant curvature ($\kp<0$, $\kp=0$, $\kp>0$). Now we have proved that these properties are also true for the PW system. The harmonic oscillator and the Kepler problem although different systems they are however endowed with very similar properties. Both are superintegrable systems possessing additional constants of motion (Fradkin tensor and Runge-Lenz vector respectively) and both can appropriately be defined on spaces of constant curvature. Now we see that a similar situation appears with the TTW system (related to the harmonic oscillator) and the PW system  (related to the Kepler problem).

  The existence of superintegrability can be studied making use of different approaches  (proof that all bounded classical trajectories are closed, Hamilton-Jacobi formalism and action-angle variables, exact solvability, degenerate quantum  energy levels, etc) but the method we have employed (complex functions with a Poisson bracket with the Hamiltonian proportional to itself) is not only useful from a practical point of view but it also reveals  interesting properties  characterizing  these two systems.

  We conclude with the following two comments. First, the Kepler problem and also the potentials $V_c$ and $V_d$ are separable in parabolic coordinates; so, it will be convenient to study the existence of a family of Kepler-related potentials  (another different linear combination with Kepler as first summand) separable in parabolic coordinates and generalizing the potentials $V_c$ and $V_d$.    Second, these system are also important at the quantum level. The properties of the functions $M_r$ and $N_\phi$ can probably be interesting  (changing functions for operators) for the study of the quantum Schr\"odinger equation by the method of factorization and ladder operators.

%-----------------------------------------------------
%%  
\section*{\bf Figures}

%----------------
\begin{figure}
\centerline{
%%  \epsfbox{QHOrphiFig1.eps}
\includegraphics{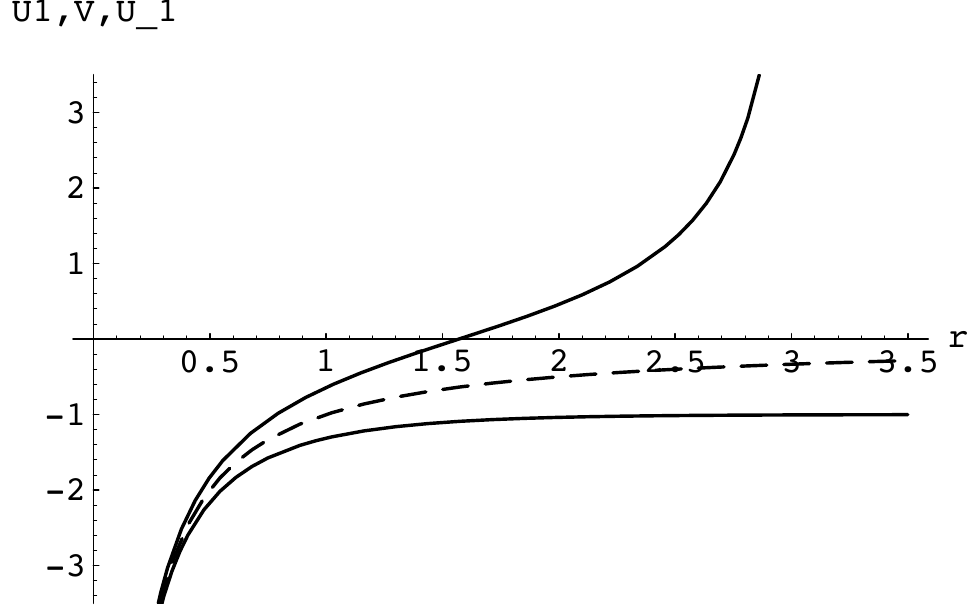}  } 

\caption{Plot of the Kepler Potential as a function of $r$, 
for the unit sphere $\kappa=1$ (upper curve), Euclidean 
plane $\kappa=0$ (dash line), and `unit` Lobachewski plane
$\kappa=-1$ (lower curve). 
The three functions are singular at $r=0$ but the Euclidean 
function $U_0=V$ appears in this formalism as making a 
separation between two different behaviours. 
In fact $U_0=V$ is the only Potential that vanish at long distances.  }
\label{Fig1}
%%\end{figure}

\end{figure}

%-----------------------------------------------------
\section*{Acknowledgments}

This work was supported by the research projects MTM--2012--33575 (MICINN, Madrid)  and DGA-E24/1 (DGA, Zaragoza). 

%----------------
 \vfill\eject
%----------------

{\small

%--------------------------------------
 }
%--------------------------------------
%----------------
\end{document}